CONDENSED MATTER

# Magnetoresistance of a Two-Dimensional Electron Gas in a Parallel Magnetic Field[1]

V. T. Dolgopolov* and A. Gold**

*Institute of Solid-State Physics, Russian Academy of Sciences, Chernogolovka, Moscow oblast, 142432 Russia
** Centre d'Elaboration de Matériaux et d'Etudes Structurales (CNRS), 31055 Toulouse, France

Received November 4, 1999; in final form, December 7, 1999

**Abstract**—The conductivity of a two-dimensional electron gas in a parallel magnetic field is calculated. We take into account the magnetic-field-induced spin-splitting, which changes the density of states, the Fermi momentum, and the screening behavior of the electron gas. For impurity scattering, we predict a positive magnetoresistance for low electron density and a negative magnetoresistance for high electron density. The theory is in qualitative agreement with recent experimental results found for Si inversion layers and Si quantum wells. © 2000 MAIK "Nauka/Interperiodica".

In recent experiments [1–9], the transport properties of the two-dimensional electron gas (2D EG) in Silicon inversion layers and GaAs heterostructures have been studied by applying a parallel magnetic field. The motivation to study transport properties came from a renewed interest into the metal-insulator transition in a 2D EG [10–12]. In the experiments, a strong positive magnetoresistance has been found in the metallic phase. The experimental fact that the magnetoresistance saturates above the magnetic field $B_c$, corresponding to a totally polarized electron system, was interpreted as a manifestation for the importance of the spin-polarization [8, 9]. At electron densities where a strong magnetoresistance is found, it was shown experimentally that the conductivity increased with decreasing temperature [5, 9, 13, 14]. This temperature dependence was successfully described by a temperature dependant screening behavior [15–17].

One expects that the transport properties of the metallic phase of a 2D EG, depending on temperature and magnetic field, are explained in the frame of a single theory. For a weakly disordered EG, we propose in the present paper an explanation of the large magnetoresistance in the metallic phase due to the magnetic field induced changes of the screening properties of the 2D EG. The corresponding temperature dependence is also described. The effect of the parallel magnetic field is to provide the spin-polarization of the EG. In the fully polarized system, the spin degeneracy is lifted and the Fermi energy increases by a factor two together with a reduction of the density of states by a factor two compared to the two-dimensional EG in zero magnetic field. In fact, we shall show that these ingredients are already sufficient to describe a positive magnetoresistance at low and intermediate electron density and a negative magnetoresistance at high electron density.

We use a minimal model in order to describe the effects of the parallel magnetic field. The term parallel means that the magnetic field is in the plane of the EG. First, we assume that the two-dimensional EG has zero width in the direction perpendicular to the interface. Second, we consider only charged impurity scattering without spin-flip processes. Screening effects are taken into account within the random-phase approximation including many-body effects described by the local-field correction.

The electron density N defines the Fermi wave number $k_F$ of the 2D EG via $N = g_s g_v k_F^2/4\pi$. Here, $g_v$ and $g_s$ are the valley and the spin degeneracy factors, respectively, and $k_F$ is the Fermi wave number. The Fermi energy $\varepsilon_F = k_F^2/2m^*$ is given by the Fermi wave number and the effective mass $m^*$. For Si inversion layers and Si quantum wells, $g_v = 2$, while for GaAs/AlGaAs heterostructures the valley degeneracy factor is $g_v = 1$. For zero field, the spin degeneracy is $g_s = 2$, while for large magnetic field the degeneracy factor is given by $g_s = 1$. For intermediate fields, the system is partially spin-polarized. We assume that the disorder is due to charged impurities of density $N_i$ located in the plane of the EG. The magnetic field applied parallel to the 2D EG plane leads to a Zeeman energy $\Delta E = \pm g^*\mu_B B/2$. Here $g^*$ is the effective Landé $g$-factor. The system will be total spin-polarized if $\Delta E$ is larger than $\varepsilon_F$. This condition defines a critical magnetic field $B_c$ for complete spin-polarization and given by $B_c = 2\varepsilon_F/g^*\mu_B$.

---
[1] This article was submitted by the authors in English.

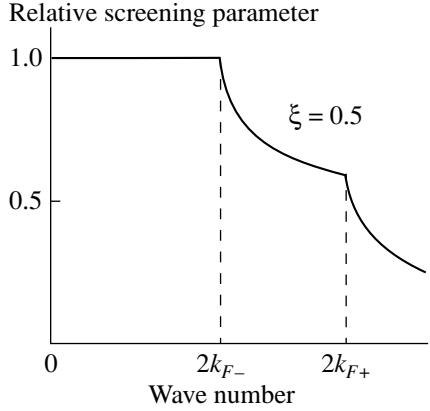

**Fig. 1.** Effective screening parameter for the partially spin-polarized 2D EG as function of the wave number.

For charged impurity scattering, the scattering time $\tau$ is given by [16, 18]

$$\frac{h}{\tau} = \frac{(2\pi)^2}{g_s g_v} \varepsilon_F \frac{N_i}{N} \frac{1}{[1 - (G(2k_F, g_s) + 2k_F/q_s)]^2}. \quad (1)$$

Here, $G(2k_F, g_s)$ is the local-field correction which describes many-body effects and depends on the spin degeneracy. Within the random-phase approximation, the local-field correction $G(2k_F, g_s) = 0$ is neglected. Most important is that the Fermi wave number $k_F = (4\pi N/g_s g_v)^{1/2}$, the screening wave number $q_s = g_s g_v/a^*$, and the Fermi energy $\varepsilon_F \propto k_F^2 \propto 1/g_s g_v$ depend on the spin degeneracy.

In fact, there are three factors which are important for the scattering time: (i) the density of states $\rho_F$ at the Fermi energy, which leads to $1/\tau \propto \rho_F \propto g_s$; (ii) the Fermi wave number (due to backscattering processes up to $2k_F$) with $k_F \propto 1/g_s^{1/2}$; and (iii) the screening wave number $q_s \propto g_s$, which enters, together with $2k_F$, the screening function $\varepsilon(q)$ and contributes as $1/\varepsilon^2(q = 2k_F) \propto 1/(1 + q_s/2k_F)^2$, which leads for $q_s \gg 2k_F$ (low density) to $[1/\varepsilon(q = 2k_F)]^2 \propto 1/g_s^3$ and for $q_s \ll 2k_F$ (high density) to $[1/\varepsilon(q = 2k_F)]^2 = 1$. Consequently, for $q_s \gg 2k_F$ the conductivity $\sigma$ is given by $\sigma \propto g_s^2$, and for $q_s \ll 2k_F$ one finds $\sigma \propto 1/g_s$. We conclude that, for $q_s \gg 2k_F$, the resistivity $\rho$ increases with increasing field $\rho(B = B_c)/\rho(B = 0) = 4$, while for $q_s \ll 2k_F$ the resistivity decreases with increasing field $\rho(B = B_c)/\rho(B = 0) = 0.5$.

We introduce the partially spin-polarized 2D EG by the densities $N_\pm$ given by $N_\pm = N(1 \pm \xi)/2$, where the spin-polarization parameter is $\xi = (N_+ - N_-)/N$ with $0 \leq \xi \leq 1$. In terms of the magnetic field, the polarization parameter is given by $\xi = g^* m_B B/2\varepsilon_F$. For the partially spin-polarized system, we argue as follows: qualitatively the spin-polarization leads to different $k_F$-vectors for spin-up $[k_{F+} = k_F(1 + \xi)^{1/2}]$ and spin-down $[k_{F-} = k_F(1 - \xi)^{1/2}]$ electrons (or holes). The effective screening parameter [18] for finite spin-polarization as a function of the wave number is shown in Fig. 1. Note that, for $k_{F-} \leq q \leq k_{F+}$, the screening parameter is strongly wave-number dependent. By taking into account this effective screening parameter for the partially polarized electron gas, we get for the conductivity in the case of $q_s \gg 2k_F$ a positive magnetoresistance

$$\frac{\sigma(0 \leq B \leq B_c)}{\sigma(B = 0)} = \frac{1-\xi}{2} + \frac{\pi(1+\xi)}{16 f(\nu)}, \quad (2)$$

with

$$f(\nu) = \arcsin(\nu)/4 + [\pi/2 - \arcsin(\nu)]/[2 - (1 - \nu^2)^{1/2}]^2, \quad (3)$$

and

$$\nu = [(1-\xi)/(1+\xi)]^{1/2}. \quad (4)$$

The function $f(\nu)$ has the correct limits $f(\nu = 0) = \pi/2$, which implies $\rho(B = 0)/\rho(B = 0) = 1$, and $f(\nu = 1) = \pi/8$, which implies $\rho(B = B_c)/\rho(B = 0) = 4$.

In Si inversion systems, the electron density corresponding to $q_s = 2k_F$ is quite high: $N = 2.8 \times 10^{13}$ cm$^{-2}$. In GaAs the density corresponding to $q_s = 2k_F$ is much lower: $N = 1.6 \times 10^{11}$ cm$^{-2}$. We obtain for $q_s \ll 2k_{F-}$ a negative magnetoresistance:

$$\frac{\sigma(0 \leq B \leq B_c)}{\sigma(B = 0)} = 1 + \xi^2. \quad (5)$$

We mention that the predicted magnetoresistance is insensitive to the angle between the electric current and the magnetic field in the 2D EG plane. This fact allows one to separate the magnetoresistance caused by spin effects from the magnetoresistance caused by orbital motion in a 2D EG with finite width. The orbital effect was discussed recently and depends on the width of the 2D EG [19].

For the 2D EG, we expect the same increase of the resistance in a magnetic field normal to the 2D EG plane if the magnetic field is smaller than the quantizing one (for instance, $B < 0.4$ T for high mobility Si inversion layers). With the magnetic field normal to the 2D EG plane, the positive magnetoresistance due to the spin will be in competition with the negative magnetoresistance due to weak localization. This means that, for a correct interpretation of the weak localization contribution, it is not enough to compare the resistance values without a field and in a weak normal field, as it is done in the literature. We suggest that spin dependent effects should be measured in a parallel magnetic field and the corresponding correction has to be introduced to describe the magnetoresistance in the normal magnetic field. We expect that the correction due to spin-polarization has the same order of magnitude as the weak localization contribution.

For the temperature dependence, using our analytical results for charged impurity scattering [16], we predict for $B = 0$ and $B \geq B_c$

$$\sigma(T < \varepsilon_F) = \sigma(T = 0)$$
$$\times \left[1 - 4\ln 2 \frac{1 - G(2k_F, g_s)}{1 - G(2k_F, g_s) + 2k_F/q_s} \frac{k_B T}{\varepsilon_F}\right]. \quad (6)$$

Note that, for $g_s = 1$, the Fermi energy increases by a factor 2 compared to $g_s = 2$. Correspondingly, the temperature dependence is weaker for a spin-polarized EG compared to an unpolarized one.

It is not straightforward to apply (1)–(6) to remote doping with a large spacer width $\alpha$, because one cannot ignore the form factor, which enters the theory [18, 20]. Remote doping is important in GaAs/Al$_x$Ga$_{1-x}$As heterostructures. For $2k_F\alpha \gg 1$ we predict that the magnetoresistance is always positive and given by $\rho(B \geq B_c)/\rho(B = 0) = 2^{1/2}$ and the linear temperature dependence of the conductivity will disappear due to a form factor, which enters the theory as $\exp(-4\alpha k_F)$ [21]

We briefly discuss the limitations of our approach. We assume a 2D EG with zero thickness in the direction normal to the plane of the EG. Our theory is formulated for weak disorder and low temperature. Therefore, (1)–(6) cannot be applied near the metal-insulator transition and the Fermi energy should exceed the temperature significantly. In silicon inversion layers, interface-roughness scattering is important for intermediate and high electron density [18]. For interface-roughness scattering, we expect the same results as found for impurity scattering.

In (2)–(5), we have ignored many-body effects described by a local-field correction. For a low electron density, where $2k_F/q_s \ll 1$, we get, in fact,

$$\frac{\sigma(B \geq B_c)}{\sigma(B = 0)} = \frac{1}{4}\frac{[1 - G(2k_F, g_s = 1)]^2}{[1 - G(2k_F, g_s = 2)]^2}. \quad (7)$$

Numerical results concerning $G(2k_F, g_s)$ for $g_s = 2$ and $g_s = 1$ [22] indicate that the magnetoresistance might increase by using a finite local-field correction.

At the present time, detailed experimental results for the conductivity of the good metallic phase in a parallel magnetic field are missing. Below, we compare results of our calculation with some experimental data from the literature. In Fig. 2, we compare curves from (2)–(4) with experimental points obtained from the high-mobility silicon inversion 2D EG of [8]. The agreement between theory and experiment is quite good for the maximal electron density $N = 2.1 \times 10^{11}$ cm$^{-2}$. However, for an electron density of $N = 1.7 \times 10^{11}$ cm$^{-2}$, the experimental magnetoresistance exceeds the theoretical one nearly twice. We believe that at this density the sample is already close to the metal-insula-tor transition, which occurs for this sample near $N \sim 1 \times 10^{11}$ cm$^{-2}$.

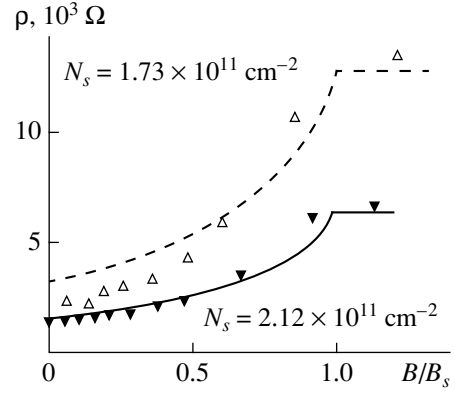

**Fig. 2.** Resistivity as function of the parallel magnetic field (in units of the critical magnetic field $B_c$ for complete spin-polarization) for parameters corresponding to silicon. The solid points are experimental results [8] for two different electron densities taken from experiments.

The ratio $\rho(B > B_c)/\rho(B)$ found in [3] for the two highest electron densities $N = 2.2 \times 10^{11}$ cm$^{-2}$ and $N = 2.6 \times 10^{11}$ cm$^{-2}$ is very close to the result of our calculation $\rho(B > B_c)/\rho(B) = 4$.

For $2k_F/q_s \ll 1$, we conclude that the coefficient of the linear temperature dependence in (6) should be $4\ln 2 = 2.77$ for $B = 0$ and $B > B_c$, independent of the magnetic field. This prediction is in good agreement with experiments [9], where $\rho(T) = \rho(T = 0)[1 + 2.9k_B T/\varepsilon_F]$ was found for Si quantum wells for $B = 0$ and $B > B_c = 9$ T. The ratio $\rho(B \geq B_c)/\rho(B = 0) = 2$, also observed in [9], might be explained by a finite spacer effect.

We emphasize that, for $2k_F/q_s \ll 1$, the coefficient of the linear temperature dependence is universal and does not depend on the local-field correction. Such a universal behavior as the function of the carrier density was already verified in experiments with GaAs heterostructures using holes [5].

The crossover point for the transition from a positive magnetoresistance to a negative magnetoresistance is given by $2k_F = q_s$ with $q_s \propto 1/a^*$ and with the corresponding carrier density $N \propto 1/a^{*2}$. With increasing mass, this density increases. Assuming that charged impurities are located in the GaAs, we expect in GaAs/Al$_x$Ga$_{1-x}$As heterostructures for $N = 3 \times 10^{11}$ cm$^{-2}$ a positive magnetoresistance for holes, as already seen in experiment [5], and a negative magnetoresistance for electrons. In conclusion, we presented a theory for the magnetoresistance of a parallel field and for the temperature dependent resistance based on the spin-polarization of the two-dimensional EG. A very important ingredient in our approach is the screening behavior of the spin-polarized system. Recent experimental results support our theoretical predictions.

This work was supported in part by the Russian Foundation for Basic Research under Grants 97-02-16829 and 98-02-16632 and the program "Nanostructures" from the Russian Ministry of Sciences under Grant 97-1024.